\documentclass[pra,onecolumn,showpacs,floatfix]{revtex4-1}
\usepackage{graphicx}
\usepackage{dcolumn}
\usepackage{bm}
\usepackage{amsmath}
\usepackage{amssymb}
\usepackage{revsymb4-1}
\begin{document}

\title{Gap solitons in optical lattices embedded into nonlocal media}
\author{YuanYao Lin$^1$, Chandroth P. Jisha$^1$, Ching-Jen Jeng$^1$,
Ray-Kuang Lee$^1$, and Boris A. Malomed$^2$}
\affiliation{$^1$Institute of Photonics Technologies, National Tsing-Hua University,
Hsinchu 300, Taiwan\\
$^2$Department of Physical Electronics, School of Electrical Engineering,
Faculty of Engineering, Tel Aviv University, Tel Aviv 69978, Israel.}

\begin{abstract}
We analyze the existence, stability, and mobility of gap solitons
(GSs) in a periodic photonic structure built into a nonlocal
self-defocusing medium. Counter-intuitively, the GSs are supported
even by a highly nonlocal nonlinearity, which makes the system
quasi-linear. Unlike local models, the variational approximation
(VA) predicts the GSs in a good agreement with numerical findings,
due to the suppression of undulating tails of the solitons.
\end{abstract}

\maketitle

\section{Introduction}

\label{Sec:intro} Solitons are self-guided wave packets propagating in
nonlinear media, maintaining the self-trapped shape. In particular, optical
solitons are supported by the balance between the material nonlinearity and
diffraction in the spatial domain or dispersion in the temporal domain \cite%
{book}. As concerns spatial solitons in planar waveguides, it is well known
that the self-focusing Kerr nonlinearity supports bright ones, while a
defocusing nonlinearity admits dark solitons. Even in the absence of
integrability, solitons readily feature quasi-particle collisions, with the
outcome depending on the relative phase between them. These properties
suggest to use solitons in various applications to all-optical
data-processing schemes and telecommunication systems \cite{soliton-ph}.

Efficient control of the transmission and localization of light may be
provided by photonic crystals (PhCs), built as structures with periodic
modulation of the refractive index (RI). They open the ways to tailor the
dispersion, diffraction, and routing of electromagnetic waves \cite%
{Joannopoulos}. Nonlinear PhCs, composed of appropriate materials, have
revealed a wealth of nonlinear optical phenomena, including the
self-trapping of localized modes in the form of the gap solitons (GSs) \cite%
{Sterke94, Mingaleev, npc-book}. These modes can be formed in self-focusing
and defocusing media alike, due to the possibility of the change in the sign
of the effective dispersion{/}diffraction in PhCs. \cite{Ostrovskaya03}.
Experimentally, GSs were first created in the temporal domain, as solitons
in a short piece of a fiber Bragg grating \cite{Krug}. Technologies based on
the use of reconfigurable (photoinduced) lattices, that have been
implemented in photorefractive crystals \cite{Efremidis2002} and nematic
liquid crystals \cite{Peccianti2002}, offer new ways to control GSs in the
spatial domain, by varying the lattice depth and spacings.

Combining the benefits of PhCs and solitons, GSs have considerable potential
for the use in photonics. GSs of matter waves have also been theoretically
studied \cite{Konotop} and experimentally created \cite{Markus} in
Bose-Einstein condensates formed by atoms with repulsive interactions,
trapped in optical-lattice potentials. Bifurcations and stability of optical
GSs were analyzed in PhCs with the local Kerr nonlinearity \cite%
{Pelinovsky2004}. However, the limited mobility of GSs in the transverse
directions, due to their pinning to the underlying lattice potentials \cite%
{Sakaguchi}, is an obstacle to the use of GSs in switching and routing
operations \cite{Crist2003, kartashov2004}.

Recently, it has been predicted that solitons supported by a \emph{nonlocal}
nonlinearity, self-focusing or defocusing, in the combination with the
effective diffraction induced by either the total internal reflection
(ordinary solitons) \cite{Torner-gap} or bandgap spectrum \cite{GP-YY}, may
move much easier across the lattice. Nonlocal effects come to play an
important role as the characteristic correlation radius of the medium's
response function becomes comparable to the transverse width of the wave
packet \cite{Wieslaw2000}. Experimental observations of nonlocal responses
have been demonstrated in various media, including photorefractive crystals
\cite{Duree93}, nematic liquid crystals \cite{Conti03}, and thermo-optical
materials \cite{Rotschild05,Efremidis2008}. The nonlocal nonlinearity
induces new features in the wave dynamics, modifying the underlying
modulational \cite{Krolikowski04}, azimuthal \cite{Anton06}, and transverse
\cite{TI-YY} instabilities. Suppression of the collapse of multidimensional
solitons \cite{Bang02}, a change of interactions between them \cite%
{Peccianti02}, the formation of soliton bound states \cite{Torner05}, merger
of colliding solitons into a standing wave \cite{nbragg-YY}, and families of
dark-bright soliton pairs \cite{nlocal-YY} were also predicted recently.

The nonlocality is known to improve the stability of solitons due to the
diffusion mechanism of the underlying nonlinearity. In the limit of strongly
nonlocal nonlinearity, the system become an effectively linear one \cite%
{Snyder97}. In such an extreme limit, the existence of GSs (for the
defocusing sign of the nonlinearity) is questionable. In this work, we
identify families of bright on-site and off-site GSs in self-defocusing
nonlinear media by means of numerical methods and analytical methods. With
the infinite range of the nonlocality, we demonstrate the existence of
spatial GSs with a finite beam's width. The analytical consideration is
based on the variational approximation (VA) with a Gaussian ansatz, similar
to how it was applied to the matter-wave GSs in Refs. \cite{VA1, VA2}.
Unlike the case of the local defocusing nonlinearity, in the nonlocal modes
the Gaussian ansatz works well not only deep inside of the bandgap, but also
close to its edge. The stability and mobility of the GS families in the
nonlocal medium are investigated too.

\section{The model and numerical results for gap solitons}

\label{Sec:Ngs}
\begin{figure}[h]
\centering
\includegraphics[width=8cm]{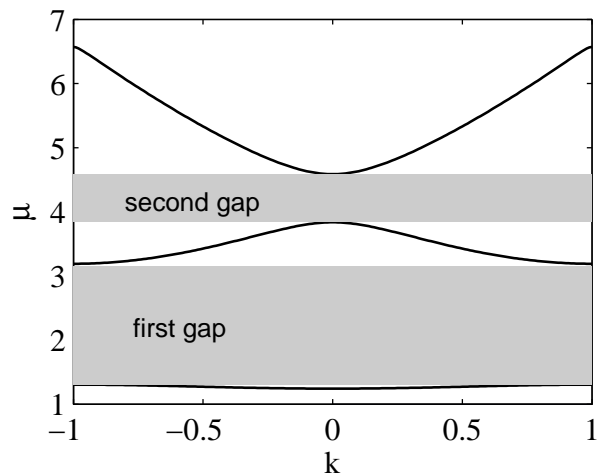}
\caption{A typical example of the spectrum, with quasi-wavenumber $k$,
induced by the linearize version of Eq. (\protect\ref{eq:phi}), with $%
V_{0}=4 $. Shaded areas are covered by the bandgaps.}
\label{Fig:F1}
\end{figure}

We consider a wave packet propagating along axis $z$ in a PhC
structure embedded into with a medium with the self-defocusing cubic
nonlocal nonlinearity. A model widely adopted for the description of such
media is \cite{Kr2001,Krolikowski04}
\begin{eqnarray}
&&i\frac{\partial \Psi }{\partial z}=-\frac{1}{2}\frac{\partial ^{2}}{%
\partial x^{2}}\Psi +V(x)\Psi +n\Psi ,  \label{eq:GPE} \\
&&n-d\frac{\partial ^{2}}{\partial x^{2}}n=|\Psi |^{2},  \label{n}
\end{eqnarray}%
where $\Psi $ is the amplitude of the electromagnetic wave, $x$ the
transverse coordinate, $n(x,z)$ a perturbation of the local RI corresponding
to the intensity-response function with an exponential kernel, and $d$ is a
parameter which determines the degree of the nonlocality of the response.
All the physical quantities and spacial coordinates are made dimensionless by normalization procedure with respect to the input beam width, wavelength, and Kerr coefficient of the nonlinear material~\cite{book} .
The limit of $d\rightarrow \infty $ corresponds to the well-known Zakharov's
system, which is a fundamental model in plasma physics (for Langmuir waves)
and other fields \cite{Zakharov}. The PhC structure is represented by the
periodic transverse potential, $V(x)=V_{0}\sin ^{2}x$ ($x$ is normalized so
as to make the period equal to $\pi $). Stationary solutions with
propagation constant $\mu $ are sought for as $\Psi (x,z)=\mathrm{exp}\left(
-i\mu z\right) \phi (x)$, which gives rise to the stationary version of Eqs. (%
\ref{eq:GPE}) and (\ref{n}):
\begin{eqnarray}
&&\mu \phi =-\frac{1}{2}\phi _{xx}+V_{0}\mathrm{sin}^{2}(x)\phi +n\phi ,
\label{eq:phi} \\
&&n-d\frac{\partial ^{2}n}{\partial x^{2}}=|\phi |^{2}.  \label{nstat}
\end{eqnarray}%
If the nonlinearity is omitted, Eq. (\ref{eq:phi}) decouples from
$n(x)$ and becomes a linear equation, which supports Bloch-wave
solutions, $\phi(x)=f(x)\exp(i\,k\,x)$, where $k$ is the
quasi-wavenumber, and $f(x)$ is a periodic function with period $\pi
$. As an example, we take $V_{0}=4$ and display the corresponding
dispersion relation, including the three lowest three bands, in
Fig.~\ref{Fig:F1}. From this diagram, it is seen that finite
bandgaps are introduced by the periodic potential; in particular,
the first finite bandgap covers a broad interval, $1.305<\mu <3.19$.

\begin{figure}[h]
\centering
\includegraphics[width=14cm]{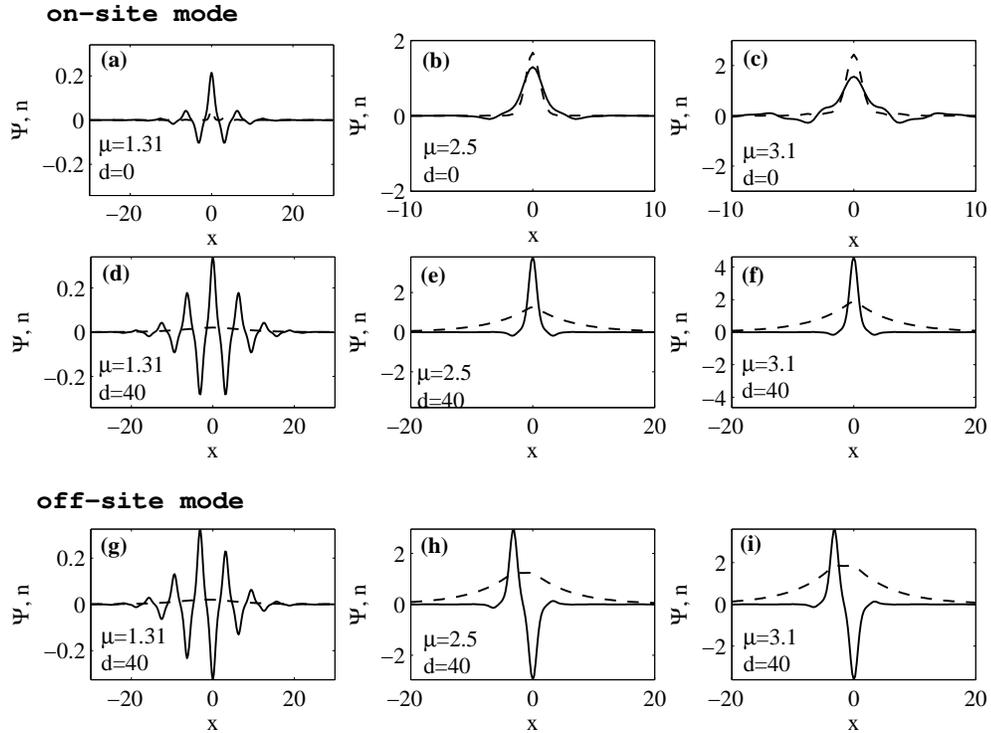}
\caption{Typical examples of gap-soliton modes in the first finite bandgap.
(a-c): On-site solutions in the local model ($d=0$). (b-f): On-site
solutions for $d=40$. (h-i): Off-site solution for $d=40$. Solid lines show
the field profiles, while the corresponding profiles of the refractive-index
perturbation, $n(x)$, are plotted by dashed lines.}
\label{Fig:F2}
\end{figure}

The nonlinearity may give rise to $x$-periodic modes \cite{dipolar-YY} or
localized GSs \cite{ypzhang2009}, with $\mu $ falling into the bandgaps,
which refers to \textit{gap solitons}. Starting with the GS solution in the
middle of the gap, we have found different families of bright solitons
numerically, using by standard relaxation technique with boundary conditions
$\phi (\pm \infty )=0$. In Fig.~\ref{Fig:F2}, we demonstrate generic
examples of the GS modes found in the first finite bandgap. Local (with $d=0$%
), and nonlocal (for $d=40$) on-site-centered GSs are shown near the bottom
of the gap in Figs.~\ref{Fig:F2}(a, d), in the middle of the gap in Figs.~%
\ref{Fig:F2}(b, e), and close to the top edge in Figs.~\ref{Fig:F2}(c, f),
with propagation constants $\mu =1.31$, $\mu =2.5$, and $\mu =3.1$,
respectively. Simplest higher-order GS solutions of the nonlocal model
(off-site-centered solitons) are presented in Fig.~\ref{Fig:F2}(g-i). The
two distinct types of the solitons, on-site and off-site, are defined by the
position of their centers with respect to the underlying periodic potential
\cite{Dabrowska,Torner-gap}.

\begin{figure}[h]
\centering
\includegraphics[width=6cm]{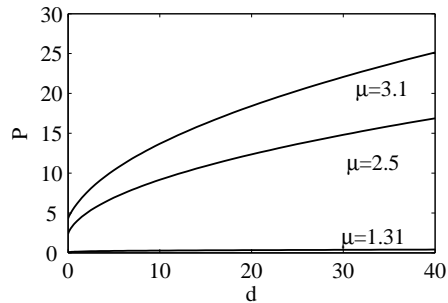}
\caption{The total power of on-site gap solitons versus the nonlocality
parameter, $d$, for three distinct values of the propagation constant taken,
respectively, near the bottom edge of the first finite bandgap's edge ($%
\protect\mu =1.31$), in the middle of the gap ($\protect\mu =2.5$), and
approaching the top band edge ($\protect\mu =3.1$).}
\label{Fig:F3}
\end{figure}

\begin{figure}[tbp]
\centering
\includegraphics[width=12cm]{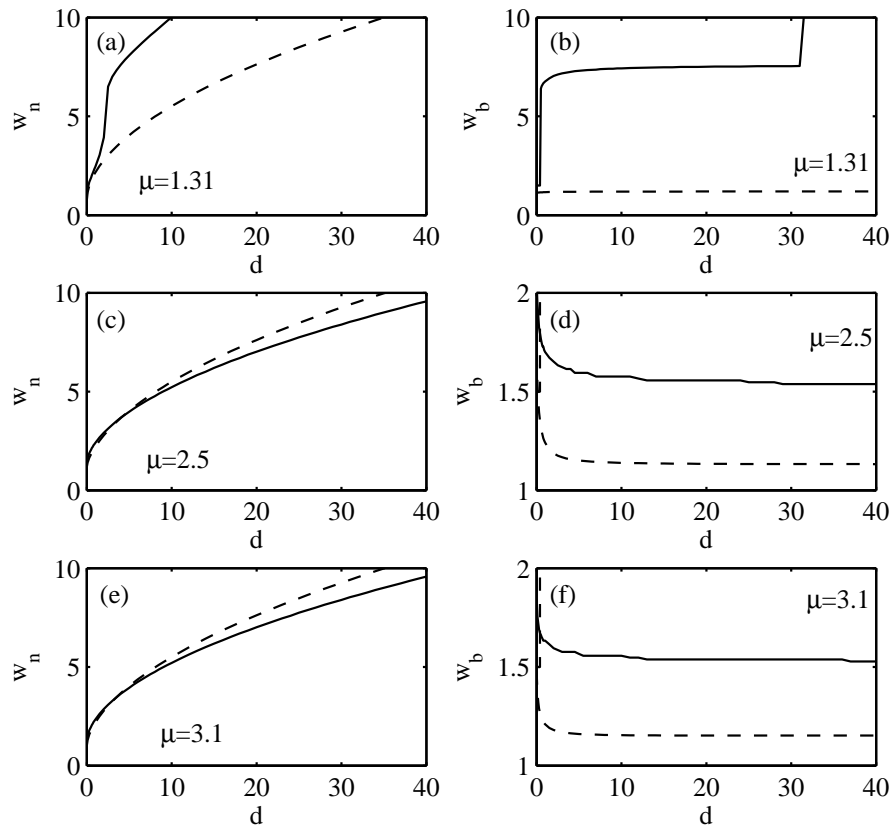}
\caption{Left column (a,c,e): The width of the refractive-index profile, $%
w_{n}$, in gap solitons, versus nonlocality parameter $d$. Right
column (b,d,f): The width of the field component of the gap
solitons, $w_{b}$, versus $d$. Numerical and variational results are
shown by solid and dashed lines, respectively. The value of the
propagation constant is fixed in each panel, as indicated.}
\label{Fig:F5}
\end{figure}

Similar to solitons in nonlocal media with the self-focusing nonlinearity
\cite{Kr2001,Krolikowski04}, the amplitude of the GSs in the present model
increases at a higher degree of the nonlocality, $d$. As a result, the
related total power, $\mathbf{P}\equiv \int_{-\infty }^{\infty }|\Psi
(x)|^{2}\,dx$, is a growing function of $d$, as shown in Fig.~\ref{Fig:F3}.
In comparison to the local nonlinear medium, with $d=0$ [see Figs. \ref%
{Fig:F2}(a-c)], the width of the RI perturbation, $w_{n}$, becomes broader
with the increase of $d$, for the focusing \cite{Kr2001,Krolikowski04} and
defocusing signs of nonlinearity alike, due to the diffusion type of the
nonlocal response. Relations between $w_{n}$ and $d$ are shown in the first
column of Fig.~\ref{Fig:F5}. On the contrary to the solitons in
self-focusing nonlocal media \cite{Torner-gap, GP-YY}, the beam's width of
the GSs in the present case, $w_{b}$, \emph{decreases} with the increase of $%
d$, as shown in the second column of Fig.~\ref{Fig:F5}. At very large values
of $d$, the beam's width in the GS approaches a constant value.

\section{Variational approximation}

\label{Sec:var} 
\begin{figure}[tbp]
\centering
\includegraphics[width=8cm]{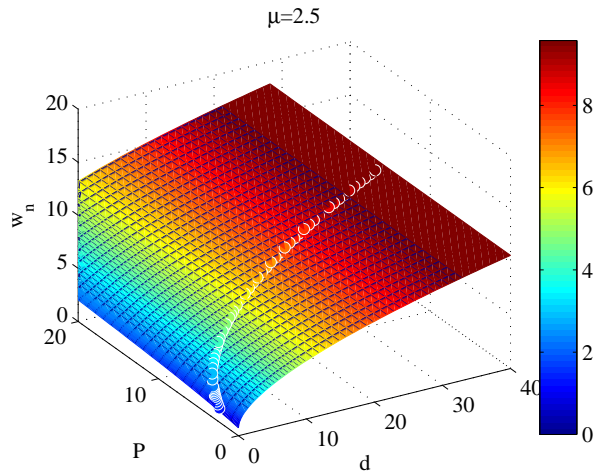}
\caption{(Color online) The surface plot for the width of the refractive- index component
of the gap soliton, $w_{n}$, at $\protect\mu =2.5$, as predicted by
variational equations~(\protect\ref{eq:vara1}), (\protect\ref{eq:vara2}).
The chain of dots represents full numerical solutions. }
\label{Fig:F4}
\end{figure}

It is well known that undulating tails in the shape of GSs,
induced by the underlying periodic potential, make the Gaussian ansatz
inappropriate as an approximation for GSs, especially close to bandgap edges
\cite{VA1}. However, Fig. \ref{Fig:F2} demonstrates that the nonlocal
nonlinearity makes GSs in the present model more localized, suggesting to
apply the VA. The Lagrangian density for Eq.~(\ref{eq:GPE}) is
\begin{equation}
{L}=\frac{i}{2}\left( \phi _{z}^{\ast }\phi -\phi _{z}\phi ^{\ast }\right)
+k|\phi |^{2}+\frac{1}{2}\left( \phi _{x}\right) ^{2}+V_{0}\mathrm{sin}%
^{2}(x)|\phi |^{2}+n|\phi |^{2}-\frac{d}{2}\left( n_{x}\right) ^{2}-\frac{%
n^{2}}{2},  \label{eq:Lagn}
\end{equation}%
Following the above argument, we adopt the Gaussian ansatz for field $\phi $
and RI perturbation $n$,
\begin{equation}
\phi (x,z)=A(z)\mathrm{exp}\left[ -\frac{x^{2}}{2w_{b}^{2}(z)}+ib(z)x^{2}%
\right] ,\quad n(x,z)=C(z)\mathrm{exp}\left[ -\frac{x^{2}}{2w_{n}^{2}(z)}%
\right] ,
\end{equation}%
with $A(z)$, $b(z)$ and $w_{b}(z)$ standing for the amplitude, chirp and
width of the field component of the GS, while $C(z)$ and $w_{n}(z)$ are the
amplitude and width of its RI counterpart. Substituting the ansatz into
Lagrangian density (\ref{eq:Lagn}) and performing the standard calculations
\cite{Progress}, we arrive at VA-generated relations between the parameters,
\begin{eqnarray}
&&\frac{1}{2w_{b}^{2}}=\frac{\left( 3d+2w_{n}^{2}\right) ^{2}}{8w_{n}^{4}\left(2w_{n}^{2}-d\right)^2}+%
\frac{\mathbf{P}\left( 3d+2w_{n}^{2}\right) }{2w_{n}\left(
d+2w_{n}^{2}\right) ^{3}}-\frac{2w_{n}\left( 3d+2w_{n}^{2}\right) }{\left(
d+2w_{n}^{2}\right) ^{2}}-\left( \mu +V_{0}\right) ,  \label{eq:vara1} \\
&&w_{b}^{2}=\frac{2w_{n}^{2}\left( 2w_{n}^{2}-d\right) }{3d+2w_{n}^{2}}.
\label{eq:vara2}
\end{eqnarray}%
Using Eqs.~(\ref{eq:vara1}) and (\ref{eq:vara2}), in Fig.~\ref{Fig:F4} we
draw the surface plot for the width of the RI profile, $w_{n}$, as a
function of total power $\mathbf{P}$ and nonlocality parameter $d$, at a
fixed value of $\mu $, and compare it to the numerical results. As expected,
the VA produces good results for the GSs taken in the middle of bandgap --
for instance, at $\mu =2.5$.

In addition, using the power and propagation constant found numerically in
Sec. \ref{Sec:Ngs}, we show in Fig. \ref{Fig:F5} that both numerical and
variation solutions represent the same trend for the widths of both
components of the GSs. In particular, for $\mu =2.5$ and $\mu =3.1$, as
shown in Fig.~\ref{Fig:F5}(c-f), $w_{n}$ increases as the square root of the
nonlocality strength, $d$, in agreement with Ref. \cite{Snyder97}. On the
contrary, $w_{b}$ drops to a finite value as $d$ increases. However, for $
\mu =1.31$, which is very close to the edge of the first finite bandgap, the
trend is completely different. This difference is explained by the known
fact that GSs with the propagation constant taken very close to edges of
bandgaps are similar to the linear Bloch waves in the linear lattice, as
seen in Fig.~\ref{Fig:F2}(d). When the amplitude of the respective
undulating tails in the GS shape is comparable to its main peak, the
Gaussian ansatz definitely fails. Nevertheless, close to the top edge of the
first finite bandgap -- for instance, at $\mu =3.1$, the ansatz still works
well because the major peak in the GS profile remains much higher than the
undulating tails in the entire nonlocality regime. Thus, the applicability
condition for the Gaussian-based VA in the system with the self-defocusing
nonlocal nonlinearity is clear: It is usable as long as the GS propagation
constant is taken not too close to the bottom edge of the first finite
bandgap.

\begin{figure}[h]
\centering
\includegraphics[width=8cm]{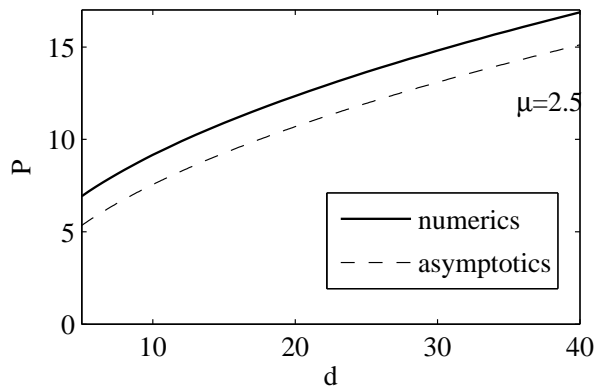}
\caption{The threshold power necessary for the formation of the gap
solitons versus the nonlocality parameter, $d$. The solid and dashed
lines depict, respectively, the numerical results and asymptotic
approximation (\protect \ref{threshold}).} \label{Fig:as}
\end{figure}

In the extremely nonlocal regime, the field component in the GSs is much
narrower than the RI profile. In this case, the RI profile may be
approximated as $n(x)\approx R(x)\equiv e^{-|x|/\sqrt{d}}/(2\sqrt{d})$,
where $R(x)$ is the RI response function. The corresponding width of the RI
profile is $w_{n}\approx 2\sqrt{d} \ln{2}$. Then, using the
quasi-linear limit similar to that developed in Ref. \cite{Snyder97}, one
can predict the threshold power necessary for the formation of the GS,
\begin{equation}
\mathbf{P}_{\mathrm{thr}}=2(\mu -\mu _{0})\sqrt{d},  \label{threshold}
\end{equation}%
where $\mu _{0}$ as the propagation constant at the edge of the first finite
bandgap. The comparison to the numerical results in this extremely nonlocal
regime is shown in Fig.~\ref{Fig:as}.

\section{Stability and mobility of the gap solitons}

Having constructed the family of the GS solutions, we analyze their
stability in the usual way, considering perturbed GS solutions as
\begin{eqnarray}
&&u=u_{0}(x)e^{ibz}+\epsilon \lbrack p(x)e^{i\delta z}+q(x)e^{-i\delta
^{\ast }z}]e^{ibz},  \label{EqSLu} \\
&&n=n_{0}+\Delta n,  \label{EqSLn}
\end{eqnarray}%
where $\epsilon \ll 1$ is the perturbation amplitude , $u_{0}(x)$ is the
unperturbed solution, and $\mathrm{Im}\{\delta \}$ is the growth rate of the
perturbations. Although the strengthening nonlocality makes the GS shape
sharper  and stronger localized, we have found, somewhat
counter-intuitively, that the on-site GS family is still stable, while its
off-site counterpart is not, cf. \cite{Pelinovsky2004}. Figure~\ref%
{Fig:MI-PN}(a) demonstrates that the nonlocality significantly reduces the
growth rate of the unstable perturbation mode for off-site solitons, as in
the case in the self-focusing nonlinearity \cite{GP-YY}. Due to its
diffusion character, the nonlocality smoothes down the undulating tails in
the RI profile, $n(x)$. It is this smoothness that stabilizes GSs, for
either sign of the nonlinearity. Therefore, as the strength of the
nonlocality increases, the GS solutions become more stable, through the
broadening of the effective potential.

\begin{figure}[h]
\centering
\includegraphics[width=12cm]{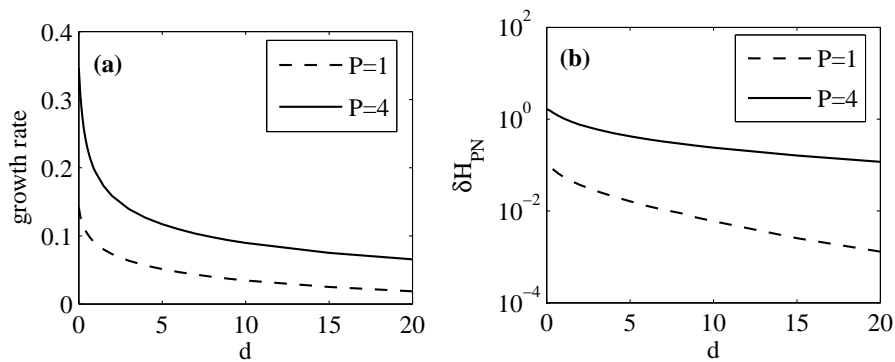}
\caption{(a) The instability growth rate and (b) the PN energy barrier
versus the nonlocality strength, $d,$ for different values of the total
power, $\mathbf{P}$.}
\label{Fig:MI-PN}
\end{figure}

The mobility of the GSs in the present model was studied through calculating
the respective Peierls-Nabarro (PN) potential barrier, which is defined as
the height of an effective periodic potential generated by the underlying
lattice. The potential barrier determines the minimum energy needed to move
the center of mass of a localized wave packet by one lattice site \cite%
{Kivshar93}. As usual, we can calculate the PN  barrier as the difference of
values of the model's Hamiltonian ($H$) between on-site and off-site modes
\cite{Torner-gap}, i.e.,
\begin{equation}
\delta H=H_{\mathrm{even}}-H_{\mathrm{odd}},
\end{equation}
\begin{equation}
H\equiv \int_{-\infty }^{\infty }\left[ -\frac{1}{2}\left\vert {\frac{%
\partial {u}}{\partial x}}\right\vert ^{2}-\frac{1}{2}\left\vert
u\right\vert ^{2}n-V_{x}|u|^{2}\right] dx.
\end{equation}%
As seen in Fig.~\ref{Fig:MI-PN}(b), in the first finite bandgap the PN
barrier is reduced in comparison to the case of the local nonlinearity, $d=0$%
, which is a natural manifestation of the nonlocality.

\section{Conclusion}

We have reported the analysis of the existence, stability, and mobility of
one-dimensional GSs (gap solitons) in the periodic potential structure
combined with the self-defocusing nonlocal nonlinearity. We have found that
the GSs become tighter localized in space, with a higher formation-power
threshold. The results have been obtained in the numerical form and
reproduced, with a reasonable accuracy, by the VA (variational approach).
Using the linear-stability analysis and calculating the PN (Peierls-Nabarro)
potential barrier, we have demonstrated that the GSs become not only more
stable, but also more mobile, with the increase of the nonlocality. The
comparison with the limit of the extreme nonlocality was reported too.
Taking into regard the possibilities offered by the currently available
technology for fabricating nonlocal nonlinear media with controllable
properties, such as photorefractive crystals, nematic liquid crystals and
thermo-optical materials, the results reported in this work may suggest new
possibilities for the design of soliton-based photonic devices. 
For instance, in photorefractive materials like SBN or LiNbO$_3$, photorefractive gratings~\cite{book} and self-defocusing nonlinearity~\cite{zhu2006} can be easily achieved by exteranl electrical field properly applied acorss the crystal axis. Also in liquid-filled photonic crystal fibers~\cite{rosberg2007}, 2D band gap effect with thermal-type self-defocusing nonlinearity is demonstrated.
It may also
be interesting to extend the model and the analysis of GSs in it (including
vortex solitons) to the two-dimensional geometry.

\section*{Acknowledgement}

This work is partly supported by the National Science Council of Taiwan with
contrasts NSC 95-2112-M-007-058-MY3, NSC 95-2120-M-001-006 and NSC
98-2112-M-007-012.

\end{document}